\documentclass[preprint,aps,showpacs,pre]{revtex4}
\usepackage[dvips]{graphicx}
\usepackage{amssymb}

\begin{document}
\title{Immittance response of an electrolytic cell in the presence of adsorption, generation,  and recombination of ions}
\author{Jamile Lorena de Paula, Jos\'e Adauto da Cruz, Ervin Kaminski Lenzi, and Luiz Roberto Evangelista\footnote{Corresponding author: Phone: +55 44 3011 5979, FAX: +55 44 3263 4623, e-mail: lre@dfi.uem.br}}
\affiliation{Departamento di F\'{\i}sica, Universidade Estadual de Maring\'a,
Avenida Colombo 5790, 87020-900 Maring\'a, Paran\'a, Brazil.}
\date{\today}

\begin{abstract}
Effects of the adsorption--desorption process on the immittance response of an electrolytic cell are theoretically investigated in the framework of the diffusional Poisson--Nernst--Planck (PNP) continuum model,  when the generation and recombination of ions is taken into account. The analysis is carried out by searching solutions for the drift--diffusion equation coupled to the Poisson's equation  relating the effective electric field to the net charge density. The effect of different ion mobilities on the immittance, i.e., situations with equal and different diffusion coefficients for  positive and negative ions, are considered. A general exact expression for the admittance in the context of the linear approximation is obtained.
\end{abstract}
\pacs{82.45.-h,77.22,-d,66.10.Ed}
\maketitle

\section{Introduction}
An important mechanism to explain the role of the ions on the electrical impedance response or immittance response  of an insulated material (solid, liquid or gel) may be the selective adsorption of ions at the solid electrodes limiting the medium~\cite{Anca,Libro}. Thus, to build a more complete model to evaluate the electrical impedance of this system it may also be necessary to take into account the dissociation of neutral particles in ionic products, the recombination of these ions giving rise
to neutral particles~\cite{mac1,mac2,mac3,mac4,mac5,mac6,mac7,mac8,mac9,mac10,mac76,mac12,mac11}, and the adsorption--desorption process occurring
at the interfaces~\cite{Barbero1,Barbero2,Batalioto1}. In a recent paper~\cite{Derfel2}, the  importance of the dissociation-association process in the determination of the electrical impedance was analyzed in detail, but without taking into account the adsorption phenomena, in the framework of the Poisson-Nernst-Planck (PNP) model. A conspicuous
effect of this dissociation-association process is the appearance  of a new plateau in the real part of the electrical impedance of the cell in the dc limit.

In this paper,  we investigate the importance of the dissociation-association phenomenon for the electrical impedance of an insulating medium containing ions~\cite{Derfel1} when the adsorption--desorption process is taken into account.
Specific adsorption effects at the electrodes have been already considered by adopting Chang-Jaff\'e boundary conditions when the dissociation-association phenomenon was present~\cite{mac76,mac12}.
 The adsorption phenomenon considered here is governed by a kinetic equation describing a chemical reaction of first kind at the interface liquid medium -- electrodes, by imposing the conservation of the mumber of particles in entire bulk plus electrodes system~\cite{memory,memory1}.  The mathematical problem to be solved is then represented by a set of fundamental equations that are different from the ones treated in Ref.~\cite{mac12}.
The strategy adopted here firstly analyze the situation for which the positive and negative ions have identical diffusion coefficients, i.e., equal mobilities. This scenario may be  useful for describing the diffusion of mobile charges in isotropic liquids in general. Subsequently, the fundamental equations are faced by means of analytical methods, and the solutions are found for the case in which the diffusion coefficients of the positive and neutral particles are zero, a scenario eventually more appropriated to the description of e.g., an insulating gel.
The results we have obtained here are however quite general.
In fact, they have been obtained in the framework of the PNP continuum model, in which the  diffusion equation is solved together with the Poisson's equation in order to establish the correct spatial profile of the electrical potential inside the sample. 

The outline of the paper is as follows. In Sec.~II, we present the fundamental equations of the problem relevant to the drift--diffusion problem for the ions in an insulating medium (solid, liquid, or gel), in the presence of the generation-recombination phenomenon, described as a first order chemical reaction, and incorporating a kinetic equation at the interface. The case in which the diffusion coefficients of the positive and negative ions are equal is considered in Sec.~III.  The particular case in which the positive and neutral particles are stuck on the polymer chains forming the gel is discussed in Sec.~III, where it is shown that the adsorption--desorption phenomenon is responsible for a low frequency plateau. In addition, the generation-recombination process may induce a plateau for intermediated frequencies depending on the thickness $\kappa\tau$ and on the Debye's length $\lambda$ considered.  Some concluding remarks are drawn in
Sec.~IV.

\section{Fundamental Equations}
\label{Fundamental}

The fundamental equations established in this Section are relevant to a sample of an insulating medium limited by two identical plane-parallel electrodes, a distance $d$ apart.  The medium, of dielectric constant $\varepsilon$, contains impurities that can decompose according to the chemical reaction $A\rightleftarrows
 B^++C^-$, where $A$ indicates the neutral specie, and $B^+$ and $C^-$ the positive and negative (monovalent) ions created in the decomposition of $A$. The dissociation constant is indicated by $k_d$, and the association constant by $k_a$.  The geometry of the sample is such that the electrodes are located at the positions $z=\pm d/2$ of an axis that is normal to the surfaces of these electrodes.
In addition to the situation treated in Ref.~\cite{Derfel2}, where the sample has the same geometry, we incorporate the adsorption--desorption process on the surface of the electrode.

If we indicate by $N_n$, $N_p$ and $N_m$ the bulk densities of neutral, positive, and negative particles, respectively, in the presence of an electric field, of electrical potential $V$,  the respective bulk densities of current of the particles are

\begin{eqnarray}
\label{eq2}
j_{n}&=&-{\cal D}_n \frac{\partial }{\partial z}N_n,  \\
\label{eq2-1}
j_p&=&-{\cal D}_p\left[\frac{\partial}{\partial z}N_{p}+ \frac{q}{k_{B}T}N_p\frac{\partial}{\partial z}V_{}\right], \quad {\rm and}\\
\label{eq2-3}
j_m&=&-{\cal D}_m\left[\frac{\partial}{\partial z}N_{m}- \frac{q}{k_{B}T}N_m\frac{\partial}{\partial z}V_{}\right],
\end{eqnarray}
in which the corresponding diffusion coefficients are ${\cal D}_n$, ${\cal D}_p$ and ${\cal D}_m$.  The equations of continuity, stating the conservation of the particles, are

\begin{eqnarray}
\label{eq1}
\frac{\partial}{\partial t}N_{n} &=& -\frac{\partial}{\partial z}j_{n} - k_{d}N_{n}+k_{a}N_{p}N_{m}, \\
\label{eq1-1}
\frac{\partial}{\partial t}N_{p} &=& -\frac{\partial}{\partial z}j_{p} +k_{d}N_{n}-k_{a}N_{p}N_{m}, \quad {\rm and}\\
\label{eq1-2}
\frac{\partial}{\partial t}N_{m} &=& -\frac{\partial}{\partial z}j_{m} +k_{d}N_{n}-k_{a}N_{p}N_{m}.
 \end{eqnarray}
The remaining equation of the model is the equation of Poisson, relating the effective electric field to the net charge density. It is obtained by considering that the electrical displacement is given by $\nabla \cdot {\bf{D}} = \rho/\epsilon$, where $\rho$ is the bulk density of the charges in the medium. This quantity is given by $\rho(z,t) = q (N_p-N_m)$, where $q$ is the ion electrical charge. In the one-dimensional case we are considering, the equation of Poisson can be simply written as

\begin{equation}
\label{eq0}
\frac{\partial^{2}}{\partial z^{2}}V(z,t) = -\frac{q}{\varepsilon}\left(N_{p}-N_{m}\right)\;.
\end{equation}
The set of fundamental equations~(\ref{eq1})~--~(\ref{eq0}) have to be solved with the boundary conditions

\begin{eqnarray}
\label{e2}
j_n(\pm d/2,t)&=&0,\\
\label{eq}
\!\!\!\!j_{\alpha}(z,t)\Bigl|_{z=\pm\,\frac{d}{2}}\!\!&=&\!\!\pm\frac{d}{dt}\overline{\sigma}_{\alpha}(t), \\
V(\pm d/2,t)&=&V_0(\pm d/2,t),
\end{eqnarray}
and
\begin{equation}
\label{Kinetic}
\frac{d}{dt}\overline{\sigma}_{\alpha}(t) = \kappa\,N_{\alpha}\!\left(\!\pm\,\frac{d}{2},t\!\right)
\!-\frac{1}{\tau}\overline{\sigma}_{\alpha}(t)
\end{equation}
related to the assumption of the adsorption--desorption process and the presence of the external voltage $V_0(t)$, where $\alpha=p$ refers to the positive ion and $\alpha=m$ to the negative ion. As already mentioned, a similar problem was worked out in a different way, by considering the Chang-Jaff\'e boundary conditions to take into account specific adsorption at the electrodes~\cite{mac76,mac12}. However, it is worth mentioning again that the equations to be solved here are different from the ones considered in Ref.~\cite{mac12} due to the kind of particles considered in both approaches. In the kinetic equation Eq.~(\ref{Kinetic}),  $\kappa$ and $\tau$ are parameters describing the adsorption
phenomenon. This equation simply states that the time
variation of the surface density of adsorbed particles $\overline{\sigma}_{\alpha}(t)$ (i.e., of both signs) depends on
the bulk density of particles just in front of the adsorbing
surface, and on the surface density of particles already adsorbed. The parameter $\tau$ has obviously the dimension of time, whereas
$\kappa$ has the dimension of a length/time. Consequently, if the adsorption
phenomenon is present, from the
kinetic equation, it follows
that there is an intrinsic thickness $ \kappa \tau$. As a final remark, we notice that by means of the Eqs.~(\ref{eq1}),~(\ref{eq1-1}), and~(\ref{eq1-2}) the generation and recombination of the ions are taken into account, whereas by means of the Eqs.~(\ref{eq}) and ~(\ref{Kinetic}), the adsorption--desorption process is incorporated to the dynamics of the distribution of the ions in the cell.

Following the developments performed in Refs.~\cite{Barbero1,Derfel2}, we write $N_n={\cal{N}}_n+n_n$, $N_p=\overline{{\cal{N}}}+n_p$, and $N_m=\overline{{\cal N}}+n_m$. We assume that the variations in the bulk density of ions due to the action of the external field are very small with respect to the values in thermodynamical equilibrium. This linear approximation implies that the external difference of potential applied to the cell is such that ${\cal{N}}_{n}\gg n_{n}$,  $\overline{{\cal{N}}}\gg n_p$ and $\overline{{\cal{N}}}\gg n_m$. Likewise, for the adsorbed quantities ($\overline{\sigma}_{p}$ and $\overline{\sigma}_{m}$), we have $\overline{\sigma}_{p}=\sigma+s_{p}$ and
$\overline{\sigma}_{m}=\sigma+s_{m}$,  with $\sigma\gg s_{p}$ and $\sigma\gg s_{m}$. Note that
${\cal{N}}_{0}={\cal{N}}+{\cal{N}}_{n}$ and $k_{d}{\cal{N}}_{n}=k_{a}\overline{{\cal{N}}}^{2}$,  with $\overline{{\cal{N}}}={\cal{N}}/(1+2\kappa\tau/d)$ and  $\overline{\sigma}=[(\kappa\tau/d)/(1+2\kappa\tau/d)]{\cal{N}}d$. By accomplishing these approximations, the condition stating the conservation of the number of particles takes the form
\begin{equation}
\label{eq3-2-1}
\left(s_{p}+s_{m}\right)+\int_{-d/2}^{d/2}\left(n_n+\frac{n_p+n_{m}}{2}\right)\,dz=0.
\end{equation}
We also consider the applied potential given by $V_0(\pm d/2,t)=\pm(V_0/2)\,\exp(i \omega t)$, where $V_0$ is the amplitude and $\omega$ the circular frequency of the applied voltage. In the limit of small applied voltage (linear limit) the problem was solved in the absence of adsorption--desorption process~\cite{mac2,Derfel2} and the solution without generation and recombination of ions was found~\cite{Barbero1,Barbero2,Batalioto1}.
In the next section, we obtain analytical solutions for the problem in some particular situations, in order to get closed expressions for the impedance and admittance of the system.

\section{Solutions, Impedance and Admittance}

In general terms, it is possible to look for solutions of the problem in the form

\begin{eqnarray}
\label{ansatz}
n_{\alpha}(z,t) &=& \eta_{\alpha} (z) e^{i \omega t}, \quad {\rm with} \quad \alpha =n, p, m,\nonumber \\
s_{\alpha}(t) &=& \sigma_{\alpha} e^{i \omega t},  \quad {\rm and} \quad \nonumber \\
V(z, t) &=& \phi(z) e^{i \omega t}.
\end{eqnarray}

Using the approximation discussed above and the ansatz~(\ref{ansatz}), the  fundamental equations of the problem can be rewritten as

\begin{eqnarray}
\label{eqx1}i \omega \eta_n&=&{\cal{{\cal D}}}_n\frac{d^2}{dz^2}\eta_n-k_d \eta_n+k_a {\cal N}(\eta_p+\eta_m)\\
\label{eqx1-1}
i \omega \eta_p&=&{\cal{{\cal D}}}_p\frac{d^2}{dz^2}\eta_p+\frac{q {\cal{N}}}{k_BT} {\cal{{\cal D}}}_p\frac{d^2}{dz^2}\phi+k_d \eta_n \nonumber \\ &-&k_a {\cal N}(\eta_p+\eta_m)\\
\label{eqx1-2}
i \omega \eta_m&=&{\cal{{\cal D}}}_m\frac{d^2}{dz^2}\eta_m-\frac{q {\cal{N}}}{k_BT}{\cal{{\cal D}}}_m \frac{d^2}{dz^2}\phi+k_d \eta_n\nonumber \\ &-&k_a {\cal N}(\eta_p+\eta_m)\\
\label{eqx1-3}
\frac{d^2}{dz^2}\phi&=&-\frac{q}{\varepsilon}(\eta_p-\eta_m) \;.
\end{eqnarray}
Equations~(\ref{eqx1})-(\ref{eqx1-3}) have to be solved with the boundary conditions, at $z=\pm d/2$,

\begin{eqnarray}
\label{eqx2}{\cal{{\cal D}}}_n\,\frac{d}{dz}\eta_n&=&0\\
\label{eqx2-1}
-{\cal{{\cal D}}}_p\frac{d}{dz}\eta_p-\frac{q {\cal{N}}}{k_BT}{\cal{{\cal D}}}_p \frac{d}{dz}\phi&=&\pm i\omega\sigma_{p}\\
\label{eqx2-2}
-{\cal{{\cal D}}}_m\frac{d}{dz}\eta_m+\frac{q{\cal{N}}}{k_BT} {\cal{{\cal D}}}_m\frac{d}{dz}\phi&=&\pm i\omega\sigma_{m}\\
\label{eqx2-3}
\phi(\pm d/2)&=&\pm V_0/2.
\end{eqnarray}
with $ \sigma_{\alpha}=\kappa\tau/(1+i\omega\tau)\eta_{\alpha}$.
The solution of the problem in the absence of adsorption-desorption process has been presented in~\cite{mac2,Derfel1}. In the following, we discuss the particular cases where ${\cal{D}}_p={\cal{D}}_m={\cal{D}}$ with ${\cal{D}}_n\neq {\cal{D}}$, and ${\cal{D}}_p={\cal{D}}_n=0$ with ${\cal{D}}_{m}={\cal{D}}\neq 0$. The first case is rather simple, and could be used to describe systems like a water solution of KCl, close to the saturation. The second case could be of some importance in describing the behavior of gels doped with salt because only negative ions contribute to the conduction mechanism, since the positive ones are stuck on the
polymer chains~\cite{galliani}.

Let us analyze the first situation which corresponds to all diffusion coefficients different from zero. By substituting in Eqs.~(\ref{eqx1}) - (\ref{eqx1-3}) ${\cal{D}}_p={\cal{D}}_m={\cal{D}}$ ${\mbox{and}}$ ${\cal{D}}_n\neq {\cal{D}}$, it is possible, after some calculation, to obtain the equation

\begin{eqnarray}
\label{eq4}
\psi''- \left(2\frac{{\cal{N}}q^2}{\varepsilon k_{B}T}\right)\psi=i\frac{\omega}{\cal{D}}\, \psi,
\end{eqnarray}
subjected to the boundary condition, at $z=\pm d/2$,
\begin{eqnarray}
-{\cal{D}}\left[\psi'+ \left(2q \frac{{\cal{N}}}{k_{B}T}\right)\phi'\right]=\pm i \omega \sigma'_{m}\;,
\end{eqnarray}
where $\psi=\eta_p-\eta_m$ and $\sigma'_{m}=\sigma_{p}-\sigma_{m}$, with $ \sigma'_{m}=\kappa\tau/(1+i\omega\tau)\psi$. These equations lead us to obtain the impedance presented in Ref.~\cite{Libro}. This feature implies the generation and recombination of ions has no effect on the impedance
when the mobility of the positive ions is equal to the mobility of the negative ions. Similar result was found in Ref.~\cite{Derfel1} in absence of
adsorption-desorption process.

Now, we address our attention to the case $D_{p}={\cal D}_{n}=0$, with ${\cal D}_{m}={\cal D}\neq 0$.
For this case, the set of fundamental equations of the model, namely
Eqs.~(\ref{eq1})~--~(\ref{eq0}),   becomes

\begin{eqnarray}
\label{eqz1}i \omega \eta_n&=&-k_d \eta_n+k_a \overline{{\cal N}}(\eta_p+\eta_m),\\
\label{eqz1-1}
i \omega \eta_p&=&k_d \eta_n-k_a \overline{{\cal N}}(\eta_p+\eta_m),\\
\label{eqz1-2}
i \omega \eta_m&=&{\cal D}\frac{d^2}{dz^2}\eta_m-\frac{q \overline{{\cal N}}}{k_BT}{\cal D}\frac{d^2}{dz^2}\phi +k_d \eta_n\nonumber \\ &-&k_a \overline{{\cal N}}(\eta_p+\eta_m), \quad {\rm and}\\
\label{eqz1-3}
\frac{d^2}{dz^2}\phi&=&-\frac{q}{\varepsilon}(\eta_p-\eta_m).
\end{eqnarray}

The boundary conditions on $\eta_n$ and $\eta_p$ are identically satisfied because ${\cal D}_n={\cal D}_p=0$ imply that the bulk densities of currents for the two types of particles vanish. The remaining boundary conditions, taking into account the surface adsorption,  are

\begin{eqnarray}
\label{v-1}
-{\cal D} \frac{d}{dz}\eta_{m}- \frac{q \overline{{\cal{N}}}}{k_{B}T}{\cal D}\frac{d}{dz}\phi &=&\pm i \omega \sigma _{m} \quad{\rm and}\quad \nonumber \\
\phi(\pm d/2)&=&\pm V_0/2,
\end{eqnarray}
with
\begin{eqnarray}
\sigma_{m}=\frac{\kappa\tau}{\left(1+i\omega\tau\right)} \eta_{m},
\end{eqnarray}
at $z=\pm d/2$.
From Eqs.~(\ref{eqz1}) and~(\ref{eqz1-1}),  it follows that $\eta_n+\eta_p=0$, and
\begin{equation}
\label{vv}\eta_{p}=-\frac{\overline{{\cal{N}}}k_{a}}{(k_{d}+\overline{{\cal{N}}}k_{a}+i\omega)}\,\eta_{m},
\end{equation}
for the spatial parts of the variations of the bulk densities of the neutral and positive particles. Equations~(\ref{eqz1-2}) and~(\ref{eqz1-3}) now become

\begin{eqnarray}
\label{eq1d1}
\!\!\frac{d^2}{dz^2}\eta_{m}\!-\! \frac{q \overline{{\cal{N}}}}{k_{B}T}\frac{d^2}{dz^2}\phi\!
+\!\frac{k_{d}}{{\cal D}}\eta_{n}\!-\!\frac{k_{a}\overline{{\cal{N}}}}{{\cal D}}\!\left(\eta_{p}+\eta_{m}\right)=i\frac{\omega}{{\cal D}} \eta_{m}
\end{eqnarray}
and

\begin{eqnarray}
\label{y}\frac{d^2}{dz^2} \phi&=&\frac{q}{\varepsilon}G\, \eta_m
\end{eqnarray}
where
\begin{eqnarray}
G=\frac{i\omega +k_{d}+2k_{a}\overline{{\cal{N}}}}{i\omega +k_{d}+k_{a}\overline{{\cal{N}}}} 
\end{eqnarray}
respectively. The solutions, taking into account the boundary conditions~(\ref{v-1}), are

\begin{eqnarray}
\eta_{m}(z)={A}\sinh (\beta z)
\end{eqnarray}
and

\begin{eqnarray}
\phi(z)=\frac{q}{\varepsilon\beta^{2}}G\eta_{m}(z)+C\,z,
\end{eqnarray}
with

\begin{eqnarray}
\beta^{2}=i\frac{\omega}{{\cal D}}+\frac{G}{\overline{\lambda}^{2}}
+\frac{i\omega k_{a}\overline{{\cal{N}}}}{\left(i\omega +k_{d}+k_{a}\overline{{\cal{N}}}\right){\cal D}}\; .
\end{eqnarray}

Boundary conditions~(\ref{v-1}) yield

\begin{eqnarray}
\label{m1}
A &=& \frac{V_0 q \beta{\cal{N}}/\left(2 k_{B}T\right)}{u\sinh\left(\beta d/2\right)+\left[d{\cal{E}}/\left(2\beta\right)\right]\cosh\left(\beta d/2\right)} \quad {\rm and}\\
\label{m2}
C &=& \frac{V_{0}}{2}\frac{\left({{\cal{E}}/\beta}\right)\cosh\left(\beta d/2\right)+ v\sinh\left(\beta d/2\right)}{u\sinh\left(\beta d/2\right)+\left[d{\cal{E}}/\left(2\beta\right)\right]\cosh\left(\beta d/2\right)},
\end{eqnarray}
with
\begin{eqnarray}
u=\frac{G}{\beta\overline{\lambda}^{2}}+\frac{d}{2} v\;, \quad v = \frac{i\omega\kappa\tau}{\left(1+i\omega\tau\right){\cal D}}, \quad {\cal{E}}= \beta^{2}- G/\overline{\lambda}^{2},
\end{eqnarray}
 and $\overline{\lambda}^{2}=2 \lambda^{2}$, where $\lambda = \sqrt{\varepsilon k_B T/(2 \overline{{\cal{N}}}q^2)} $ is the Debye's screening length.

From the previous results it is possible to obtain the electric field as follows
\begin{eqnarray}
E(z,t)&=&-\frac{\partial }{\partial z}V(z,t)=-\frac{d}{dz}\phi(z)e^{i\omega\,t}
\end{eqnarray}
and, consequently, from the Coulomb theorem: $E(d/2,t)=-\left(\Sigma(t)+q\sigma(t)\right)/\epsilon$,  where $\Sigma$ is the surface density of charge on the electrode at $z=d/2$ and $q\sigma=q\,(\sigma_{p}-\sigma_{m})$ is the net adsorbed charge at $z=d/2$. These quantities are relevant to obtain the admittance $Y$ of the system.
The current is determined by the equation $I=Sd\Sigma/dt$,  which lead us to the result

\begin{eqnarray}
I =\frac{i\omega \beta}{q} S {A} e^{i\omega t}\left\{\!G\cosh\left(\beta\frac{d}{2}\right)+
\frac{\kappa\tau\beta}{1+i\omega\tau}\sinh\left(\beta\frac{d}{2}\right)\!
+\frac{\varepsilon\beta}{q{A}}C\,\right\},
\end{eqnarray}
from which the admittance, $Y=I/V$, of the sample (cell) may be determined. It is given by

\begin{eqnarray}
\label{Admittance1}
Y = i\frac{\omega q S}{\beta V_{0}}{A} \left\{\!G\cosh\left(\beta\frac{d}{2}\right)+
\frac{\kappa\tau\beta}{1+i\omega\tau}\sinh\left(\beta\frac{d}{2}\right)\!
+\frac{\varepsilon\beta}{q{A}}C\,\right\}.
\end{eqnarray}
From this equation it is possible to find the impedance related to this system with adsorption--desorption in presence of
generation and recombination of ions taking the relation $Z=1/Y$ into account.

\begin{figure}[htp]
 \centering \DeclareGraphicsRule{ps}{eps}{}{*}
\includegraphics*[scale=.4,angle=0]{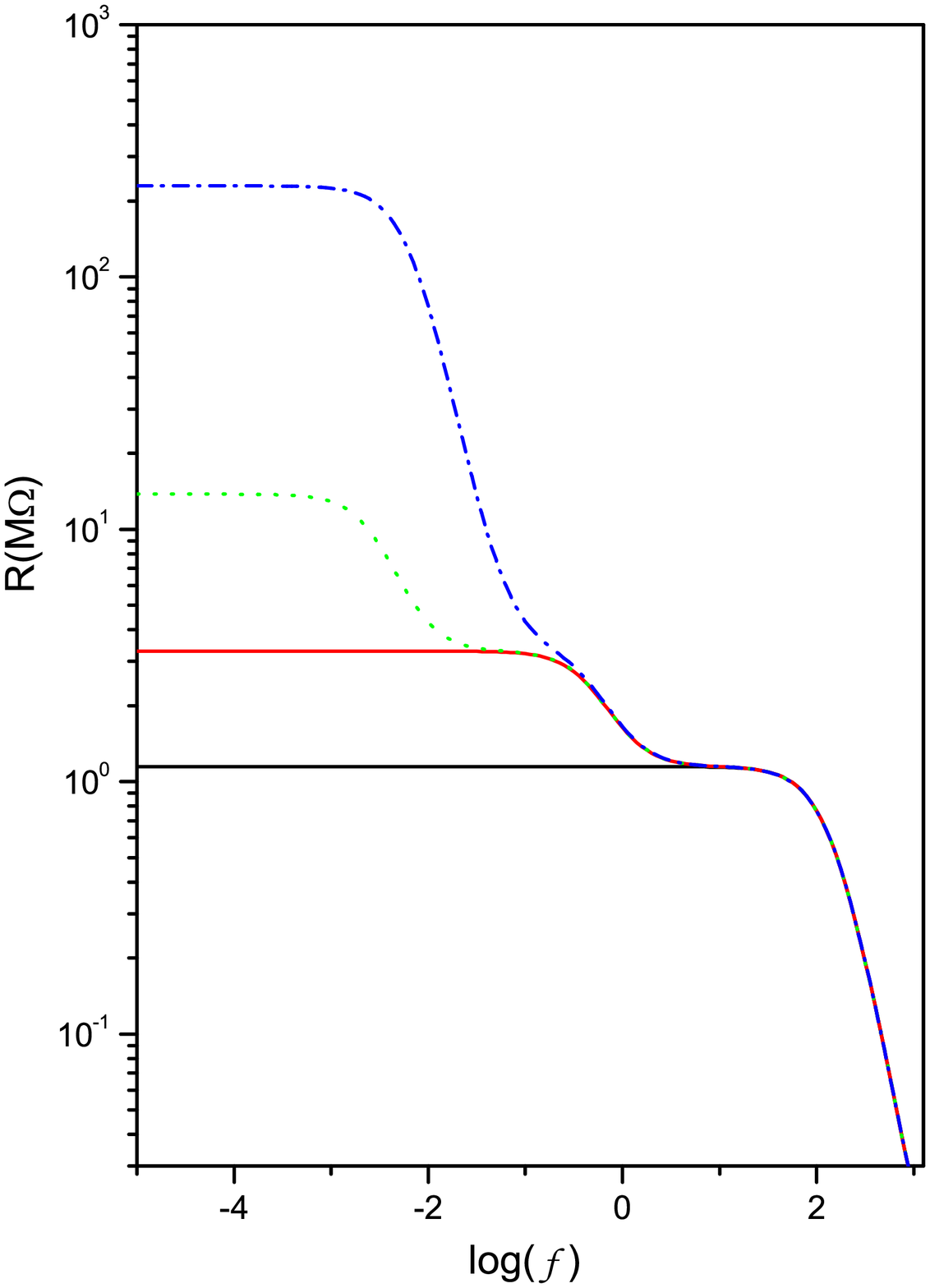}
\caption{{Real part of the electrical impedance of the cell, $R={\mbox{Re}}\,[Z]$,
versus the frequency of the applied voltage, $f = \omega/(2 \pi)$, for different values of $\kappa$, $\tau$, $k_{d}$, and $k_{a}$.}}
\label{Fig1}
\end{figure}

 Figure~\ref{Fig1} shows the behavior of the impedance for some significant situations. The black (solid) line is the case without adsorption--desorption in absence of dissociation and association, i.e., $\kappa=0$, $k_{d}=0$, and $k_{a}=0$. The red line is the case $k_{d}=0.1\; s^{-1}$,  with $k_{a}= 2\times 10^{-20}\; m^{3} s^{-1}$ in absence of the adsorption--desorption. The green (dotted) line corresponds to the case $k_{d}=0.1\; s^{-1}$,  with $k_{a}= 2\times10^{-20}\; m^{3} s^{-1}$ in presence of the adsorption--desorption process,  with $\kappa=10^{-10}ms^{-1}$ and $\tau=50s$. The blue (dashed-dotted) line corresponds to the case $k_{d}=0.1\; s^{-1}$,  with $k_{a}= 2\times10^{-20}\; m^{3} s^{-1}$, $\kappa=10^{-7}m s^{-1}$ and $\tau=50s$. One observes that the green dotted and the blue dashed-dotted lines take the effect of adsorption--desorption and association and dissociation of ions into account. They evidence that the adsorption--desorption process may play an important role at low frequency while the generation and recombination of ions has predominant effect for intermediated range of frequencies.

\begin{figure}[htp]
 \centering \DeclareGraphicsRule{ps}{eps}{}{*}
\includegraphics*[scale=.4,angle=0]{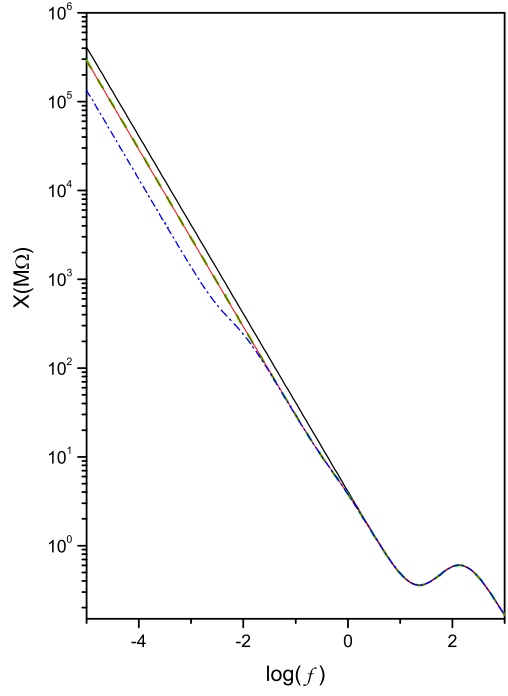}
\caption{{ Imaginary part of the electrical impedance of the cell, $X=-{\mbox {Im}}\,[Z]$,
versus  frequency of the applied voltage, $f = \omega/(2 \pi)$, for different values of $\kappa$, $\tau$, $k_{d}$, and $k_{a}$ (see the text).  }}
\label{Fig2}
\end{figure}
\begin{figure}
 \centering \DeclareGraphicsRule{ps}{eps}{}{*}
\includegraphics*[scale=.4,angle=0]{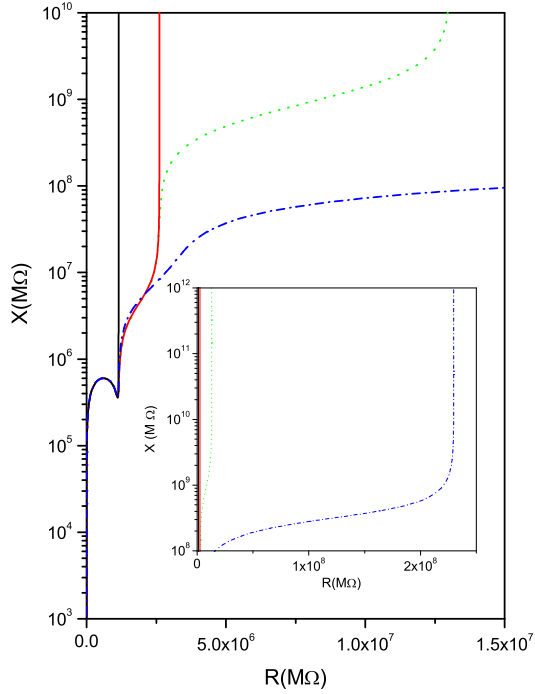}
\caption{{ Parametric plot of $X$
versus $R$ for different values of $\kappa$, $\tau$, $k_{d}$, and $k_{a}$. }}
\label{Fig3}
\end{figure}

Figure~\ref{Fig2} shows the behavior of the imaginary part of the the impedance for the same set of parameters used in Fig.~1.
Note that the imaginary part of the impedance is not sensible to the
adsorption--desorption when $\kappa\tau < \overline{\lambda}$, as shown by the green dotted line. On the other hand, for $\kappa\tau > \overline{\lambda}$ the effect of this process is evidenced (see the blue dashed-dotted line). These features can be established by analyzing  the low frequency limit of the admittance. Indeed, in the $\omega \rightarrow 0$ limit, the asymptotic behavior of the admittance is given by

\begin{eqnarray}
\label{asymptotic}
Y&\approx&i\omega\overline{Y}-(i\omega)^{2}\widetilde{Y},
\end{eqnarray}
where

\begin{eqnarray}
\label{Ad1}
\overline{Y}&\approx&\frac{S\epsilon}{\overline{\lambda}}\left[\frac{\kappa \tau}{\overline{\lambda}}+\sqrt{\frac{H}{F}}\coth\left(\frac{d}{2\overline{\lambda}}\sqrt{\frac{H}{F}} \right) \right]\; {\rm and} \\
\label{Ad2}
\widetilde{Y}&\approx&\frac{S\epsilon}{4{\cal D}
\left(\overline{\lambda}F\right)^2} \left\{dk_{a}\overline{{\cal{N}}}{\cal D}-d\lambda^{2}FH\right.
\nonumber \\&+& \left.
2\kappa\tau F^{2}\left[\left(2{\cal D}+d\kappa\right)\tau-4\overline{\lambda}^{2}\right]
+ 2\lambda\sqrt{\frac{F}{H}} \right. \nonumber \\ &\times& \left.\left[k_{a}{\cal D}\overline{{\cal{N}}}
+FH\left(2d\kappa\tau-3\overline{\lambda}^{2}\right)\right]\coth\left(\frac{d}{2\overline{\lambda}}\sqrt{\frac{H}{F}} \right)\right.
\nonumber \\ &+& \left.d\left(3FH\overline{\lambda}^{2}-k_{a}{\cal{N}}{\cal D}\right)\coth\left(\frac{d}{2\overline{\lambda}}\sqrt{\frac{H}{F}} \right)\right\},
\end{eqnarray}
with $F=k_{d}+k_{a}\overline{{\cal{N}}}$ and $H=k_{d}+2k_{a}\overline{{\cal{N}}}$. By using Eq.~(\ref{asymptotic}), it is possible to show that the asymptotic behavior of the impedance for low frequency is given by $R = {\mbox{Re}}\left(Z\right)\approx \widetilde{Y}/\overline{Y}^2$ and $X = {\mbox{Im}}\left(Z\right)\approx 1/\left(i\omega\overline{Y}\right)$.
In Fig.~\ref{Fig3}, a parametric plot of $X$ versus $R$ is reported. As the frequency increases, the effect of the adsorption--desorption is evidenced before the process of generation and recombination of ions and it may also be present in the high frequency limit.
\begin{figure}
 \centering \DeclareGraphicsRule{ps}{eps}{}{*}
\includegraphics*[scale=.4,angle=0]{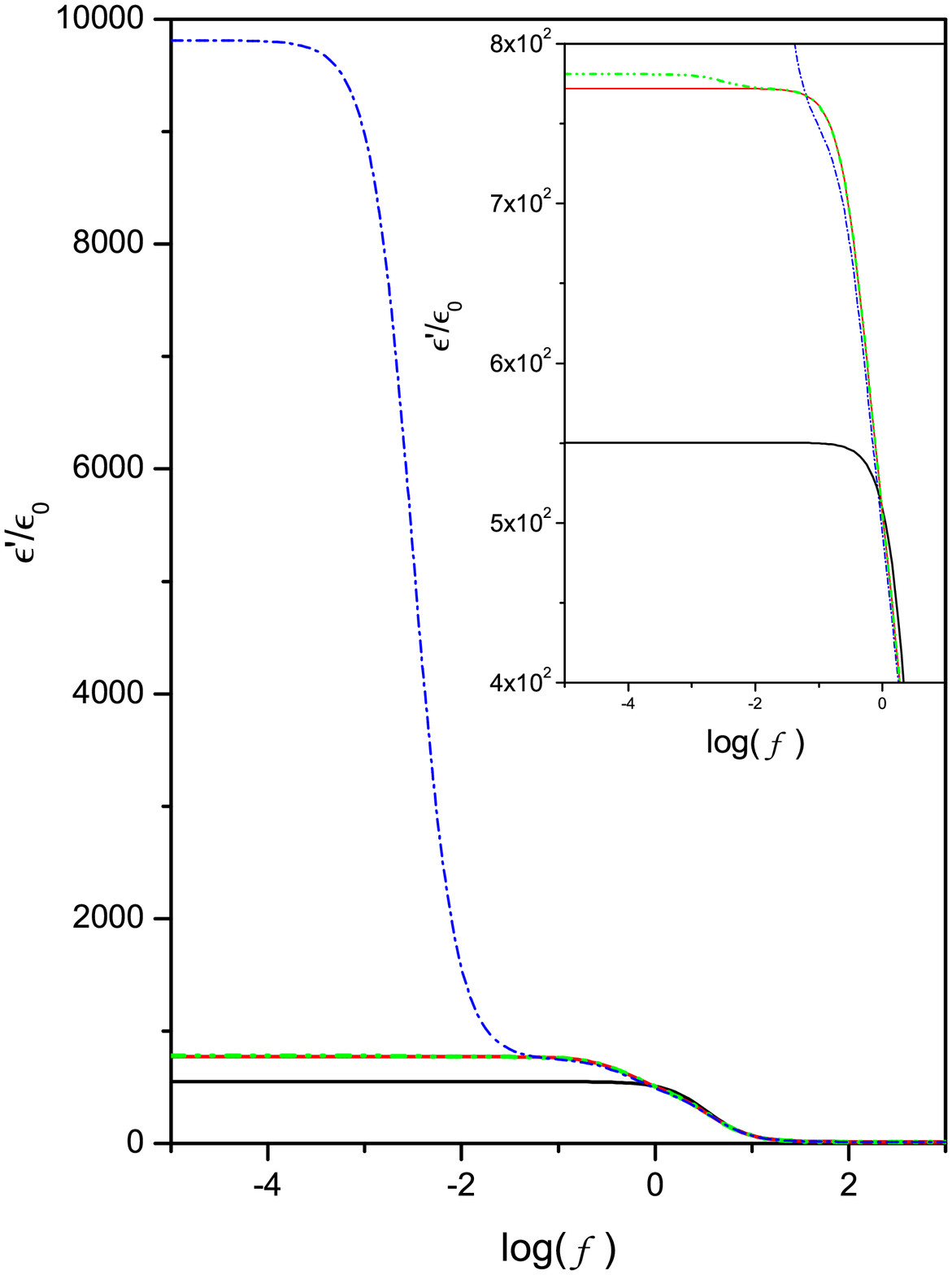}
\caption{{ $\epsilon'/\varepsilon_{0}$
versus $f = \omega/(2 \pi)$ for different values of $\kappa$, $\tau$, $k_{d}$, and $k_{a}$.}}
\label{Fig4}
\end{figure}
\begin{figure}
 \centering \DeclareGraphicsRule{ps}{eps}{}{*}
\includegraphics*[scale=.4,angle=0]{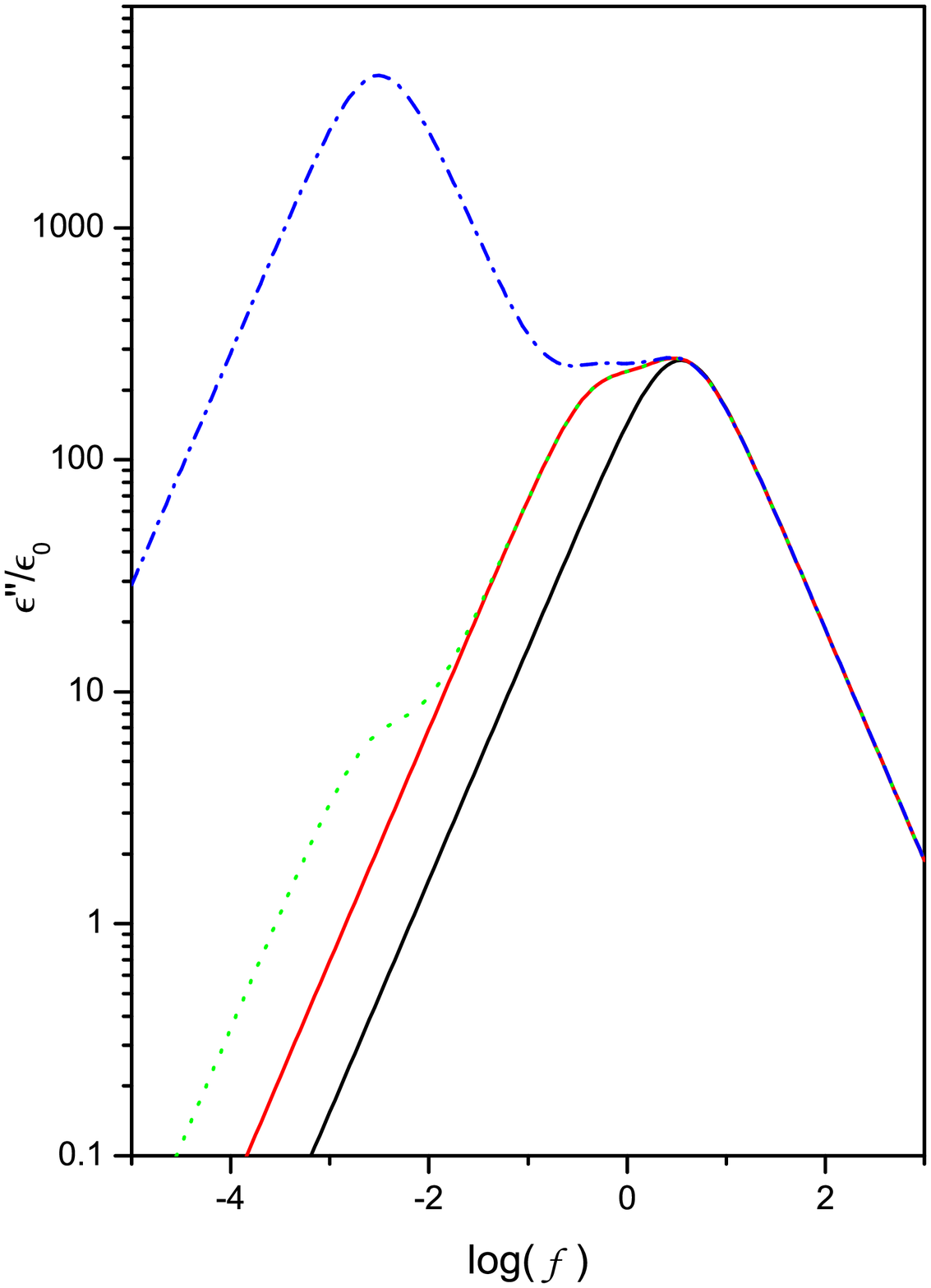}
\caption{{$\epsilon''/\varepsilon_{0}$
versus $f = \omega/(2 \pi)$ for different values of $\kappa$, $\tau$, $k_{d}$, and $k_{a}$.}}
\label{Fig5}
\end{figure}
In Figs.~\ref{Fig4} and~\ref{Fig5}, we illustrate the behavior of $\epsilon'$ and $\epsilon''$ for different situations with adsorption--desorption and generation and recombination of ions. In particular, the set of parameters used to plot these figures is the same of Fig.~\ref{Fig1}. In all these figures, we are also considering the ions monovalent with $q=1.6 \times10^{-19} A\, s$ and the temperature is such $K_{B}T/q=0.025V$. The geometric parameters of the cell are supposed to be $d=25\mu m$ and $S=2 \times10^{-4}m^{2}$. We also assume that the dielectric constant of the insulating material is $\varepsilon = 6.7\varepsilon_{0}$, the bulk density of impurities in thermodynamic equilibrium is ${\cal{N}}_{0}=10^{20} m^{-3}$, and the diffusion coefficient of the negative ions is ${\cal D} = 8.2 \times 10^{-11}m^{2}s^{-1}$.

\section{Concluding Remarks}
\label{Conclusion}
We have investigated the effect of the adsorption--desorption
process on the electrical response of an electrolytic cell (a finite-length situation) when the generation and recombination of ions are present.
We have performed this analysis by considering two situations. The first of them is characterized by equal mobilities for positive and negative ions,  which shows that the
electrical response is not sensible to the generation and recombination of ions. It is only influenced by the adsorption--desorption process. The other case considers different mobilities for the ions.
It is shown that the electrical response can be
influenced by both effects, i.e., adsorption--desorption and generation and recombination of ions. In this sense, one concludes from Fig.~\ref{Fig1} and Fig.~\ref{Fig2} that the system may exhibit different behaviors for the electrical response. In fact, in
Fig.~\ref{Fig1} the generation and recombination process governs the behavior of the system for intermediate frequencies.
Indeed, one can see the presence of initial part of a second plateau and,  after it,  the third plateau due to the adsorption--desorption process of ions (the green dotted line). The blue dashed-dotted line is essentially dominated by the adsorption--desorption process. In Fig.~\ref{Fig2}, it is shown that the imaginary part of the impedance may not manifest the effect of the adsorption--desorption process as in the real part depending on the thickness $\kappa\tau$ and the Debye length  $\lambda$ (see, for example, the green dotted and blue dashed-dotted lines). On the other hand, this effect and the generation--recombination of ions are evidenced for the blue dashed-dotted line which,  for the real part,  is essentially governed by the adsorption--desorption process. Thus, the values of the thickness $\kappa\tau$ and the Debye length $\lambda$ are useful to inform us about the influence of the surface and bulk effects on the dynamic of the ions in an electrolytic cell.
We hope that the present theoretical approach, due to its generality, may be useful to discuss  the immittance response of an electrolytic cell when both, the process of generation and recombination of ions and the adsorption--desorption phenomenon, are taken into account by means of a set of fundamental equations charactering a continuum diffusional model.
\section*{ACKNOWLEDGEMENT}
Many thanks are due to J. Ross Macdonald and G. Barbero for very useful and illuminating  discussions.
This work was partially supported by the National Institutes of Science and Technology  of Complex Systems -- INCT-SC (E. K. L.) and Complex Fluids -- INCT-FCx (L. R. E.).


\begin{thebibliography}{99}
\bibitem{Anca} G. Barbero and A. L. Alexe-Ionescu, Liquid Crystals {\bf 32}, 943 (2005).
\bibitem{Libro} G. Barbero and L. R. Evangelista,  \emph{Adsorption Phenomena and
Anchoring Energy in Nematic Liquid Crystals}, (Taylor \& Francis,
London, 2006).
\bibitem{mac1} J. R. Macdonald, Phys.\ Rev.\ {\bf 92}, 4 (1953).
\bibitem{mac2} J. R. Macdonald and Donald R. Franceschetti, J.\ Chem.\ Phys.\ {\bf 68}, 1614 (1978).
\bibitem{mac3} D. R. Franceschetti and J. R. Macdonald, J.\  Appl.\ Phys.\ {\bf 50}, 291 (1979).
\bibitem{mac4} D. R. Franceschetti and J. R. Macdonald, J.\ Electrochem.\ Soc.\ {\bf 129}, 1754 (1982).
\bibitem{mac5} J. R. Macdonald,  J.\ Electrochem.\ Soc.\ {\bf 135}, 2274 (1988).
\bibitem{mac6} J. R. Macdonald, J.\ Chem.\ Phys.\ {\bf 116}, 3401 (2002).
\bibitem{mac7} J. R. Macdonald, Phys.\ Rev.\ B {\bf 71}, 184307 (2005).
\bibitem{mac8} J. R. Macdonald, J.\ Phys.: Condens.\ Matter {\bf 17}, 4369 (2005).
\bibitem{mac9} J. R. Macdonald, J.\ Phys.: Condens.\ Matter {\bf 18}, 629 (2006).
\bibitem{mac10}J. R. Macdonald, J.\ Phys.\ Chem.\ B {\bf 112}, 13684 (2008).
\bibitem{mac76} J. R. Macdonald, J.\ Electroanal.\ Chem.\ {\bf 70}, 17 (1976).
\bibitem{mac12} J. R. Macdonald and D. R.  Franceschetti, J. Chem. Phys. {\bf 68}, 1614 (1978).
\bibitem{mac11} All the works of Macdonald are accessible, in pdf format, at  http://jrossmacdonald.com
\bibitem{Barbero1} G. Barbero, Phys. Rev. E {\bf 71}, 062201 (2005)
\bibitem{Barbero2} G. Barbero, M. Becchi, A. Strigazzi, J. LeDigabel, and A. M. Figueiredo Neto, J. App. Phys. {\bf 101}, 044102 (2007).
\bibitem{Batalioto1} F. Batalioto, O. G. Martins, A. R. Duarte, and A. M. Figueiredo Neto, Eur. Phys. J. E  {\bf 34}, 10 (2011).
\bibitem{Derfel2} G. Derfel, E. K. Lenzi, C. R. Yednak, and G. Barbero, J.\ Chem.\ Phys.\ {\bf 132}, 224901 (2010).
\bibitem{Derfel1} G. Derfel and  G. Barbero, J.\ Mol.\ Liq.\ {\bf 150}, 43 (2009).
\bibitem{memory}E. K. Lenzi, C. A. R. Yednak, and L. R. Evangelista, Phys. Rev. E {\bf 81}, 011116 (2010).
\bibitem{memory1}R. S. Zola, E. K. Lenzi, L. R. Evangelista, and G. Barbero, Phys. Rev. E {\bf 75}, 042601 (2007).
\bibitem{galliani}J. C. Gidding and H. Eyring, J. Phys. Chem {\bf 59}, 416 (1955).

\end{thebibliography}
\end{document}